\documentstyle[11pt,newpasp,twoside,epsf]{article}
\begin{document}
\title{A Polarization Flare in 3C 273: A Clue to Jet Physics}
\author{Lara L. Cross and Beverley J. Wills}
\affil{Astronomy Department, University of Texas at Austin, Austin, TX, 78712}
\author{J. H. Hough}
\affil{Department of Physical Sciences, University of Hertfordshire, Hatfield, Hertfordshire AL10 9AB}
\author{J. A. Bailey}
\affil{Anglo-Australian Observatory, P.O. Box 296, Epping, NSW 2121, Australia}
\begin{abstract}
We present $UBVRIJHK$  polarization and flux density
observations of the quasar 3C 273 obtained during a time of outburst over two
weeks in 1988 February. We have modelled these data with two 
power law components, each with wavelength-independent position angle.
These components are roughly perpendicular. The steeper-spectrum component has higher
infrared polarized flux density, with the electric vector approximately transverse to the projected direction of the VLBI jet.
The K-band polarized flux density and position angle, and possibly
the spectral index of the two components, are correlated with a time lag of less than a day. We explain our results
in terms of a shocked jet model with two nearly co-spatial components: a shock
component with magnetic field approximately perpendicular to the jet and the other with magnetic
field approximately parallel to the jet.
\end{abstract}
\section{The Data}
Our data consist of polarization position angle (E-field), percentage 
polarization, and total flux at $UBVRIJHK$ for 3C 273 for each night of UT
10, 12-18 February 1988.  The data for a representative night (12 Feb.) are
shown in Figure 1, where the solid lines represent the best-fitting model for
that night (see \S 2).  The data were taken at UKIRT with the Mark II Hatfield 
Polarimeter (Hough, Peacock, \& Bailey 1991), which makes simultaneous
optical and near-infrared observations, important for this variable blazar.
The data have been corrected for interstellar polarization and Galactic
extinction. A brief discussion of these data was presented by Wills (1991).

At the time of our observations 3C 273 was in an outburst state (Courvoisier
et al. 1988). This outburst has since been associated with the ejection of
VLBI component C9 ($t_0=1988.1\pm0.1$, B\aa \aa th et al. 1991). At this time
the polarization was higher than usual.

In our data both the percentage polarization and the position angle are very
frequency-dependent, and this dependence changes from night to night. The
frequency dependence of $\theta$, particularly a 90\deg\ flip on 14 Feb., is
suggestive of the presence of two perpendicular polarized components of 
similar amplitude but different spectral shape.

\section{Modelling}
We have modelled our data with two power laws in polarized flux density that vary  from night to night in amplitude, spectral
index, and position angle. The addition of two such 
components for 12 Feb. is illustrated in Figure 2. 
In all cases we find that the two components are approximately perpendicular,
with the E-vector of the steeper component transverse to the projected jet
direction. The model reproduces the data
well. Slight changes in the model
parameters (for example, $<$3$^{\circ}$ in position angle, $<$0.05 in spectral
index, and $<$0.05 mJy in amplitude)  produce noticeably worse fits
 to the data. The two-component models
are thus well-constrained by the data. We find that the components are
correlated in amplitude and position angle, and probably spectral index 
(significance levels of $>$99.75\%, $>$99\%, and $>$93\% respectively) with a time
lag of much less than one day. The changes of the amplitudes and position
angles of the components with time are illustrated in Figure 3.
\begin{figure}
\plotone{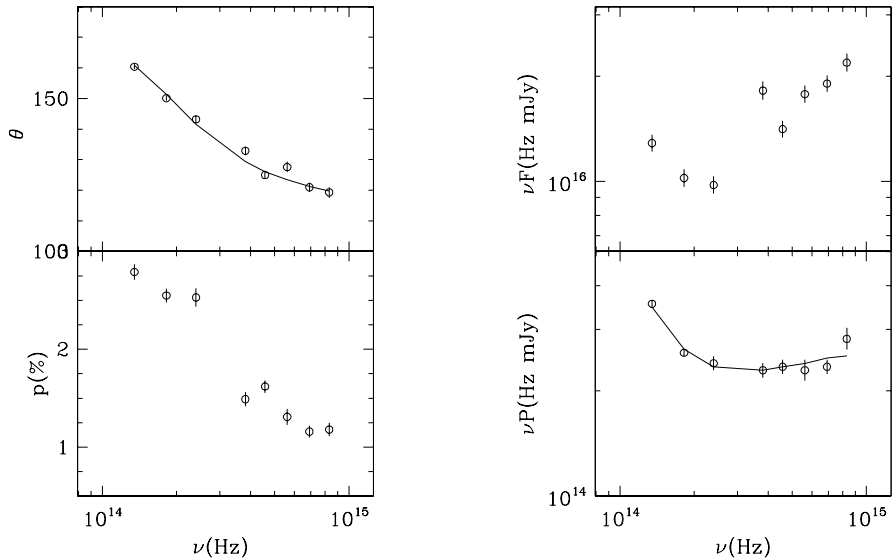}
\caption{Frequency-dependence of polarization position angle and percentage,
 total  and
polarized flux (in mJy), corrected for Galactic extinction and interstellar polarization,
 for 12 February, 1988.  Solid line represents the best-fitting
model. The high value of the total flux in I-band is due to the presence of
the H$\alpha$ line in this band.}
\end{figure}
\begin{figure}
\plotone{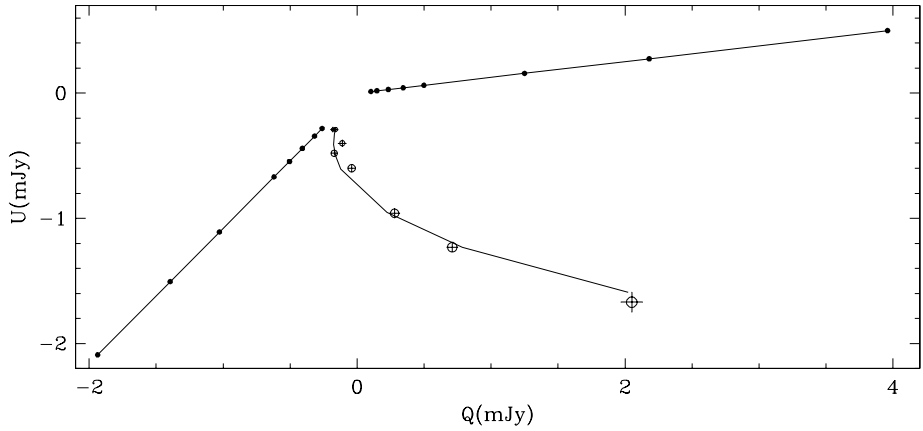}
\caption{Polarized flux, best model, and model components for 12 February, 
1988 in Stokes parameter representation. Circle size indicates wavelength,
with larger circles for longer wavelength. Filled dots represent Q and U of
the model components at the wavelengths of the passbands. The components add
vectorially to give the total model.}
\end{figure}
\begin{figure}
\plotone{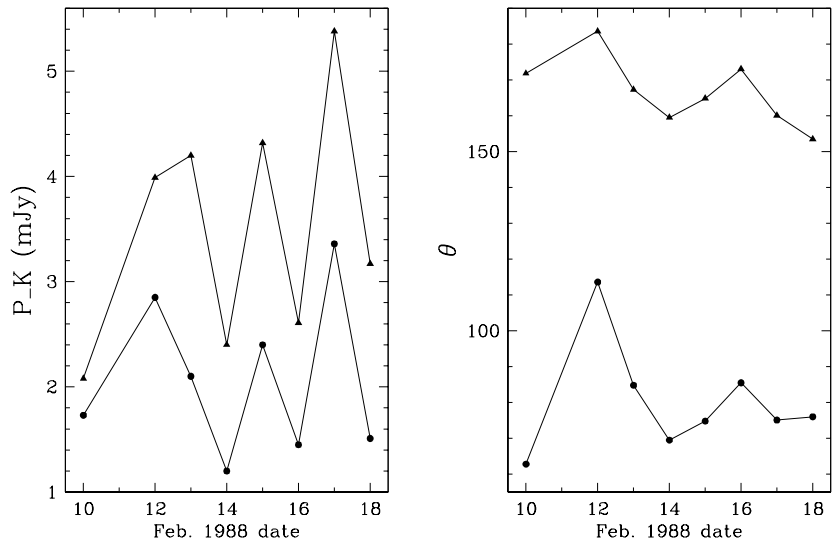}
\caption{Changes of model $K$-band amplitude and position angle for the two
model components as a function of time. Circles represent
component 1, which is approximately parallel to the VLBI jet, and triangles
represent component 2, perpendicular to the jet. The parameters for the two
components generally rise and fall together. Results are similar for the
model spectral indices. For clarity data for each component are connected by
straight lines.}
\end{figure}
 
\section{Discussion}
We explain the polarization of 3C 273 in outburst with two synchrotron
components. Since these components are nearly perpendicular they tend to cancel
each other out, resulting in low polarized flux. This blazar-like polarization
is further diluted by a strong, unpolarized non-blazar continuum, resulting in
the low observed percentage polarization for this object. 

The outburst emission of blazars is often explained in terms of a shocked jet
model like that of Marscher \& Gear (1985). Such models normally involve a
quiescent component and a shock component. We instead require two components
associated with the shock. These components must be nearly co-spatial to
account for the very short time lag between their polarizations. The simplest model would
involve an underlying magnetic field that is parallel to the jet, but which is
compressed so as to be perpendicular in the region of the shock. One of our
components, which has polarization parallel to the jet and  a flatter spectrum,
would originate from the region of the perpendicular B-field, as expected from
the Marscher \& Gear model. The other component, with perpendicular 
polarization, would be due to high energy electrons that have leaked from the
shocked region and are accelerated where the B-field is parallel to the jet.
We see these two components superimposed.

\end{document}